\documentclass[fleqn,usenatbib,useAMS]{mnras}

\usepackage{newtxtext,newtxmath}

\usepackage[T1]{fontenc}

\DeclareRobustCommand{\VAN}[3]{#2}
\let\VANthebibliography\thebibliography
\def\thebibliography{\DeclareRobustCommand{\VAN}[3]{##3}\VANthebibliography}

\usepackage{graphicx}	
\usepackage{amsmath}	
\usepackage{xcolor}

\newcommand{\noopsort}[1]{}

\title[The first JWST white dwarf debris disk]{The first white dwarf debris disk observed by \textit{JWST}}

\author[A. Swan et al.]{Andrew Swan$^{1}$\thanks{E-mail: \href{mailto:Andrew.Swan@warwick.ac.uk}{Andrew.Swan@warwick.ac.uk}},
Jay Farihi$^{2}$,
Kate Y. L. Su$^{3}$,
and Steven J. Desch$^{4}$,
\\
$^{1}$Department of Physics, University of Warwick, Coventry CV4 7AL, UK\\
$^{2}$Department of Physics \& Astronomy, University College London, London WC1E 6BT, UK\\
$^{3}$Steward Observatory, University of Arizona, Tucson AZ 85721, USA\\
$^{4}$School of Earth and Space Exploration, Arizona State University, Tempe AZ 85287, USA
}

\date{Accepted XXX. Received YYY; in original form ZZZ}

\pubyear{2023}

\begin{document}
\label{firstpage}
\pagerange{\pageref{firstpage}--\pageref{lastpage}}
\maketitle

\begin{abstract}
This letter reports the first \textit{JWST} spectroscopy of a white dwarf debris disk, giving a preliminary assessment of the salient features, and recommendations for future observations. The polluted and dusty star WD\,0145+234 experienced a major collisional event in its circumstellar disk in 2018, accompanied by an infrared outburst, and subsequently a gradual decrease in thermal emission. Time-series NIRSPEC observations demonstrate that the circumstellar disk is returning to a quiescent state with a $T\approx1000$\,K infrared excess similar to the bulk of known dusty white dwarfs. MIRI spectroscopy reveals a 9--12\,{\micron} solid-state emission feature consistent with silicate minerals as seen in debris disks observed with {\em Spitzer} IRS. The strength and morphology of the silicate feature appear unchanged relative to the continuum in spectra taken over a year apart, consistent with steady-state collisional evolution of the circumstellar debris. A tentative emission feature around 7\,{\micron} may be due to carbonates, and if confirmed would indicate aqueous alteration in the parent body.
\end{abstract}




\section{Introduction}

The gradual paradigm shift from interstellar to planetary pollution in white dwarf stars has taken well over four decades, and since the launch of the NASA {\em Spitzer Space Telescope} there has been increasing evidence that secures this scientific framework.  Over a century has passed since the first detection of heavy elements in a white dwarf (vMa\,2; \citealt{vanmaanen1917}), via photographic plate technology and visual identification of spectral lines, and now the NASA {\em James Webb Space Telescope} (\textit{JWST}) is providing an unprecedented, digital view of planetary systems that could hardly be imagined in that bygone era.

There are many significant developments that led to the current picture of planetary systems surviving and dynamically rejuvenating around white dwarfs, where the first is arguably the realisation that any primordial, atmospheric heavy elements should not persist, but instead gravitationally settle \citep{schatzman1958}.  The first robust calculations of atmospheric diffusion times (e.g.\ \citealt{fontaine1979}) confirmed the idea that the commonly detected metals in white dwarfs require an external source, and within a decade the first speculations of a planetary origin had emerged \citep{alcock1986}.

Soon thereafter came the infrared detection of a debris disk orbiting within $\sim1$\,$\textrm{R}_{\sun}$ of G29-38 \citep{zuckerman1987} at the NASA Infrared Telescope Facility, but it took another decade to detect ultraviolet and optical metal lines in this iconic polluted white dwarf \citep{koester1997}.  It would then take the better part of yet another decade to identify the second white dwarf surrounded by circumstellar dust (GD\,362; \citealt{becklin2005,kilic2005}), where those infrared observations were motivated by the presence of strong metal pollution in the host \citep{gianninas2004}.

The launch of {\em Spitzer} transformed these few dusty white dwarfs into a few dozen over the 16-yr mission lifetime, establishing the universal correlation between the warm ($T\sim1000$\,K) circumstellar dust and photospheric metals, and detecting silicate minerals in situ for all eight disks with spectroscopy (see e.g.\ review by \citealt{farihi2016}).  These infrared insights were accompanied by the pivotal realisation that the photospheric abundances of heavy elements should precisely reflect those of the planetary materials, and, with state-of-the-art diffusion models, the chemical compositions of entire extrasolar planetary bodies can thus be determined \citep{zuckerman2007,jura2014}.

This methodology has proven powerful, and is providing a window onto exoplanetary bulk chemistry and assembly that is not available for main-sequence stars.  The planetary bodies that pollute white dwarfs are dominated by the same four elements (O, Mg, Si, Fe) that comprise the terrestrial planets of the solar system, and are similarly poor in carbon \citep{klein2010,gansicke2012}.  With few exceptions, the accreted materials are refractory rich and volatile poor, and thus akin to solar system asteroids (but not comets), and water is sometimes indicated but rarely dominant \citep{farihi2013,xu2017}.  Parent body sizes are generally inferred to be analogous to large asteroids, but in some cases dwarf planet or larger masses are indicated \citep{farihi2010,hollands2018}.

More than 12 years passed between the loss of cryogen aboard {\em Spitzer} -- terminating its spectroscopic capacity -- and the launch of {\em JWST}.  During the interim, based on the post-cryogenic {\em Spitzer} and {\em WISE} missions limited only to 3.5- and 4.5-{\micron} photometry, it became increasingly clear that white dwarf debris disks are variable in flux and hence dust mass, where ongoing collisions must be responsible \citep{swan2019,swan2020}.  These empirical results imply that recurrent mass influx is necessary in white dwarf debris disk evolution, and that the former model of a quiescent, Kronian-ring analogue cannot be the whole picture \citep{wyatt2014,Kenyon2017collisions}.

Further supporting dust production over a wide range of orbital separations are the handful of polluted white dwarfs that exhibit deep, irregular, and evolving transits \citep{vanderburg2015,guidry2021}.  Despite obscuring dust clouds with scale heights of several hundred km in these systems, currently only one transiting system (WD\,1145+017) has an infrared excess detectable with {\em WISE} or {\em Spitzer} \citep{vanderbosch2021,farihi2022}.  Transiting systems and variable debris disks raise myriad questions that have remained out of reach since 2009, but now {\em JWST} can provide access to a wide range of wavelengths where many debris disks are most salient, and can probe more deeply into their wider system architecture, with unprecedented spectroscopic sensitivity.

WD\,0145+234 is a bright dusty white dwarf that experienced an exceptional collision event in 2018, accompanied by an infrared outburst and subsequent decline \citep{wang2019,Swan2021}. It is a 13\,000-K star of around 0.67\,$\text{M}_{\sun}$ that exhibits metal pollution, and its disk has a gaseous emission component \citep{McCleery2020, Melis2020gas}.  At $G=14.0$\,mag and 29.4\,pc, it is one of the brightest and nearest white dwarfs with infrared-detected dust. This letter presents the first {\em JWST} observations of a dusty white dwarf via WD\,0145+234.

\section{Observations and data reduction}

WD\,0145+234 was observed under \textit{JWST} program 1647, a multi-epoch campaign using both NIRSPEC and MIRI spectroscopy \citep{jakobsen2022,rieke2015}. Data derive from Science Data Processing Subsystem version 2022\_3a, and are reduced using Science Calibration Pipeline version 1.11.3 \citep{jwstpipeline}, using reference files defined in Calibration Reference Data System (CRDS) context \texttt{jwst\_1105.pmap}, which incorporates updates from the Cycle~1 calibration program.

The six NIRSPEC visits were all identically configured. The data were taken in bright object time-series mode, using the prism with the SUB512 array, covering the full 0.6--5.3\,{\micron} wavelength range of the instrument. Each 1-h visit consists of several hundred individual integrations of about 7\,s duration. Visits were requested on a logarithmically-increasing cadence, subject to target visibility. The first pointing occurred on 2022 July~26, and the next five followed at intervals of around 3, 6, 12, 21, and 357\,d.

The standard recipe was followed to reduce the near-infrared time-series data, with the following modifications during stage-1 processing. (1)~Snowball detection was enabled during the jump detection step. (2)~The outputs of the dark current subtraction step were modified to mitigate $1/f$ noise, estimated from the unilluminated uppermost and lowermost eight rows of the subarray. Medians of those pixels were taken for each column in each group, and subtracted. (3)~The 1D spectra were extracted using a 7-px aperture along the cross-dispersion direction, following a third-order polynomial fitted to the trace in each epoch. Small, time-dependent shifts in the trace are neglected in this first-look study, but will be important to account for in future work studying the detailed behaviour of the circumstellar material.

The two MIRI spectra were taken during the first and last visits with the Medium-Resolution Spectrometer (MRS) \citep{Wells2015}, using all three grating settings to cover the full wavelength range of 4.9--27.9\,{\micron} at $R\approx1500$–4000, with a total time on target of 3336\,s per visit. The standard MRS observation template optimised for point sources was used, including target acquisition on the source itself, a four-point spatial dither pattern, and the \texttt{FASTR1} readout pattern.

For the mid-infrared data, the reduction steps in stages~1 and~2 were run with default parameter settings, except that the residual fringe step was switched on to produce flux-calibrated images. The wavelength-dependent sensitivity loss in MRS data is accounted for by the \textit{JWST} calibration pipeline version used here, so the two epochs of MRS data are expected to be on the same flux calibration footing. These images were passed to the stage-3 pipeline to construct spectral cubes for each of the 12 MRS sub-bands, with the outlier detection step turned on to remove transient `warm' pixels during the cube-building process. The final 1D spectra were extracted using the the nominal aperture setting for point sources with the \texttt{ifu\_autocen} switch enabled, determining the target position from median-stacked images for each sub-band. Inspecting the outputs revealed that three out of 12 segments of the spectrum taken in the final visit were compromised by instrumental artefacts, so those data were excluded from the reduction. MIRI data in the longest-wavelength channel (18--28\,\micron) had low signal-to-noise (S/N), as the target is faint and dedicated background observations were not obtained, so they are not analysed further.

The full-width-at-half-maximum of the MRS PSF is wavelength-dependent \citep{argyriou23_mrs}, so the default extraction aperture radius is varied from 0\farcs53 at {4.9\,\micron} to 2\farcs1 at {29.6\,\micron}, with the corresponding sky annulus ranging from 1\farcs3 to 2\farcs0 (inner radius) and 3\farcs2 to 3\farcs9 (outer radius). The pipeline aperture sizes are chosen to be large enough to minimise the brighter-fatter effect \citep{argyriou23_brighter_fatter}. However, due to the faintness of the target, a second reduction was made with a smaller extraction aperture, aiming to reduce noise. A fixed 3-px aperture (corresponding to 0\farcs39 in channel~1) and sky annulus of 6--10\,px (0\farcs78--1\farcs3 in channel~1) were used, with aperture corrections derived using three A~stars observed in \textit{JWST} flux calibration program 1536. The smaller aperture performed better than the pipeline default: for the first and last visits, respectively, it achieved reductions of 24\,per\,cent and 48\,per\,cent in the scatter, as measured against a 0.25-{\micron} moving average in the {9--12\,\micron} region. The custom reduction is therefore adopted here.

The two instruments have overlapping wavelength coverage near 5\,{\micron}, and comparing integrated fluxes in the crossover region gives $f_{\text{NIRSPEC}}/f_{\text{MIRI}}=1.08$ for the first visit, and 1.01 for the last. The absolute spectrophotometric uncertainty measured using standard stars is larger for MIRI than for NIRSPEC \citep{Boker2023, Argyriou2023}, so the MIRI spectrum is rescaled to equalise the integrated fluxes between instruments, enabling joint analysis.

\section{Analysis}

\subsection{Time-series data}

WD\,0145+234 is not expected to exhibit pulsations, so its photospheric flux should be constant over the observatory lifetime.  This stability is exploited to mitigate the unknown systematic uncertainties in absolute flux calibration, by rescaling each NIRSPEC frame to equalise the integrated fluxes where photospheric emission dominates the spectrum.  The region below 1.1\,{\micron} is chosen for this purpose, but the \ion{Ca}{ii} triplet near 8500\,\AA~is excluded to avoid its circumstellar emission features, which are known to vary in stars of this class \citep{Melis2020gas,GentileFusillo2021GasDiscs}.  Potentially variable thermal emission from orbiting dust contributes on the order of 0.01\,per cent of the flux shortward of 1.1\,{\micron}, so this method has the potential to improve on the nominal 2~per cent absolute flux calibration accuracy targeted by the observatory. The spectra for each frame were rescaled to equalise the photospheric fluxes, where the scaling factors have a standard deviation of 0.7~per cent.

\begin{figure}
\includegraphics[width=\columnwidth]{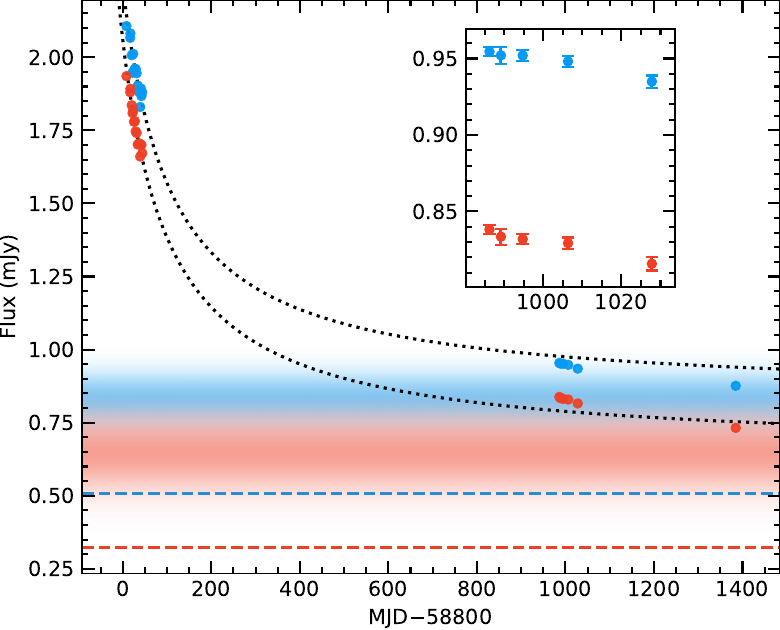}
\caption{Lightcurves showing the 3--5\,{\micron} flux of WD\,0145+234 in \textit{Spitzer} IRAC channels 1 (blue) and 2 (red). The data were taken by \textit{Spitzer} (left of centre) and \textit{JWST} (right of centre). The \textit{JWST} points are generated by integrating the NIRSPEC spectrum over the IRAC transmission profiles. The inset panel zooms in on the first five epochs of JWST data, to highlight the subtle downward trend. Coloured horizontal bands indicate the range of pre-outburst fluxes measured by the \textit{WISE} spacecraft, while the horizontal dashed lines indicate the photospheric flux. Dotted black lines show $1/t$ decay curves fitted to the data, as described in the text.}
\label{figureLightcurve}
\end{figure}

Time-series photometry was calculated by integrating the NIRSPEC spectra over the bandpasses of \textit{Spitzer} IRAC channels~1 and~2, which have effective wavelengths of 3.6 and {4.5\,\micron}, respectively.  These wavelengths probe the dust-dominated continuum emission and can be compared with prior observations.  Medians of the per-frame integrated fluxes for each epoch are used to produce the light curve shown in Figure~\ref{figureLightcurve}. Uncertainties in relative per-epoch fluxes are estimated at 0.6~per cent, but additional systematic uncertainties apply to absolute fluxes. The dust emission is less than half as bright as observed by \textit{Spitzer} 2.6\,yr previously. At that time, the {3.6-\micron} flux dropped by about 11~per cent over 36\,d, whereas the decreases measured by \textit{JWST} are 2~per cent over 42\,d during the first five epochs, and 8~per cent over the full 399-d program. This drop in flux is mirrored in the MIRI spectrum: the integrated flux at 6--{7\,\micron} (thermal continuum) and 10--{11\,\micron} (mineral features) decreases by 19~per~cent and 27~per~cent respectively. These values are consistent with expectations of less dimming at shorter wavelengths due to flux dilution from the photosphere.

Detailed analysis of infrared variability for WD\,0145+234 is beyond the scope of this letter, but a heuristic test shows that the gradually diminishing dust emission remains consistent with a collisional cascade model \citep{Swan2021}. The infrared flux from colliding debris is expected to evolve with time approximately as $f\propto1/t$ \citep{Dominik2003}. Decay curves of that form provides a reasonable fit to the data, as shown in Figure~\ref{figureLightcurve}. The fit was constrained by requiring the decay to begin after the maximum observed flux (MJD~58\,692), and before the first \textit{Spitzer} measurement shown in Figure~\ref{figureLightcurve} (MJD~58\,808). The zero point towards which the flux decays is set to the average pre-outburst flux in each channel, neglecting potential changes in dust temperature between observational epochs, which could alter the flux ratio between the two channels. By contrast, when the zero points are set to photospheric fluxes, a good fit cannot be obtained.

\subsection{NIRSPEC and MIRI spectra}
\begin{figure}
\includegraphics[width=\columnwidth]{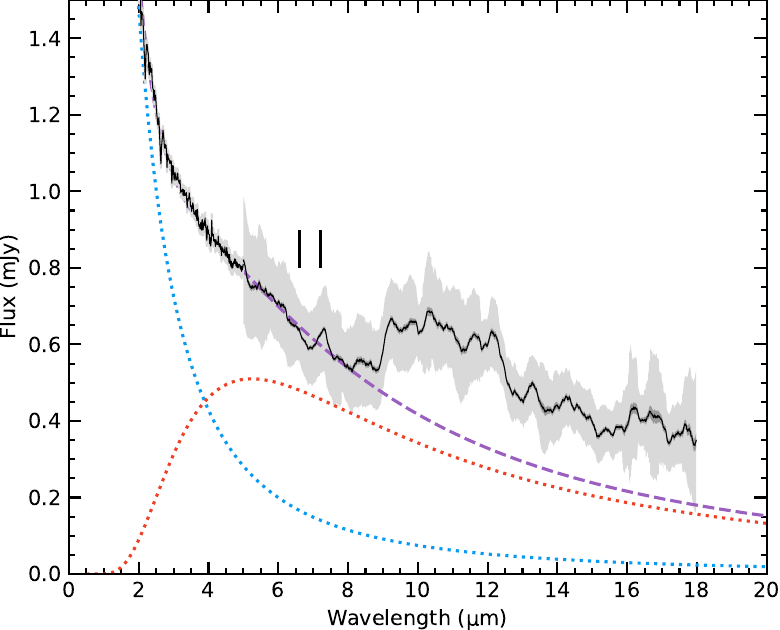}
\caption{Infrared spectrum of WD\,0145+234, showing NIRSPEC and MIRI MRS data taken during the first visit (2022 July 26). The MIRI spectrum is scaled to match NIRSPEC where their wavelength coverage overlaps, and smoothed using a 0.25-{\micron} moving average. MIRI channel~4 is not shown, due to low S/N. Light grey shading shows uncertainties per 7-s frame for NIRSPEC ($\lambda<5$\,\micron), and per resolution element for MIRI ($\lambda>5$\,\micron). Dark grey shading shows the standard error on the moving average, which is barely discernible at this scale, and almost certainly smaller than the systematic errors. The blue dotted curve shows a 13\,000\,K blackbody representing the star, the red dotted curve shows a 960\,K blackbody representing continuum emission from the dust, and the purple dashed line is their sum. Vertical bars mark the tentative carbonates feature discussed in the text, and shown in detail in Figure~\ref{figureCarbonates}.}
\label{figureFullspectrum}
\end{figure}

Figure~\ref{figureFullspectrum} shows the spectra taken during the first visit. The \textit{JWST} calibration program is ongoing, and the data reduction pipeline remains under active development, so full analyses of the spectral energy distribution and its variability are reserved to future work. However, the thermal continuum can be closely approximated by two blackbodies: one representing the star (13\,000\,K; \citealt{McCleery2020}), and the other the circumstellar dust.

While simplistic, blackbody models have proved useful for characterising dusty systems (e.g.~\citealt{Rocchetto2015, Dennihy2017}). The infrared stellar spectrum is in the Rayleigh-Jeans regime, and thus largely insensitive to temperature. The historically popular flat opaque disk model \citep{Jura2003} is an unsuitable choice for the the circumstellar material, as it cannot account for the variable flux that characterises this system, and fails entirely when applied to another star in the same class \citep{Swan2020dust, GentileFusillo2021GasDiscs}. However, whatever the configuration of the dust, its warmest component likely dominates the emission, which can thus be captured to first order by a blackbody.

The model is fitted to the NIRSPEC data at each epoch, where the free parameters are the dust temperature and the two scaling factors. The NIRSPEC data are then supplemented by MIRI data shortward of {8\,\micron} where available, and the fit is repeated. Results are shown in Table~\ref{tableFits}. The dust temperature remains stable during the first five epochs, but appears to increase slightly in the sixth. The inclusion of MIRI data reduces the uncertainties sufficiently that -- under this simple model -- the increase between epochs 1 and 6 is significant at the 5-$\upsigma$ level.

\begin{table}
\caption{Temperatures and fractional luminosities ($\tau$) for circumstellar dust at each epoch, derived from the two-blackbody fits described in the text. Uncertainties are purely statistical, but permit comparison between epochs.}
\label{tableFits}
\begin{center}
\begin{tabular}{lrlll}
\hline
Epoch &    MJD & $\tau\,(\%)$ & $T_{\textrm{dust},\lambda<5}$\,(K) & $T_{\textrm{dust},\lambda<8}$\,(K) \\
\hline
1     &  59786 &     $0.42$ &                     $962\pm15$ &                      $973\pm4$ \\
2     &  59789 &     $0.42$ &                     $964\pm15$ &                             -- \\
3     &  59794 &     $0.41$ &                     $964\pm15$ &                             -- \\
4     &  59806 &     $0.41$ &                     $961\pm15$ &                             -- \\
5     &  59827 &     $0.40$ &                     $960\pm15$ &                             -- \\
6     &  60184 &     $0.34$ &                     $984\pm18$ &                     $1010\pm5$ \\
\hline
\end{tabular}
\end{center}
\end{table}

\begin{figure}
\includegraphics[width=\columnwidth]{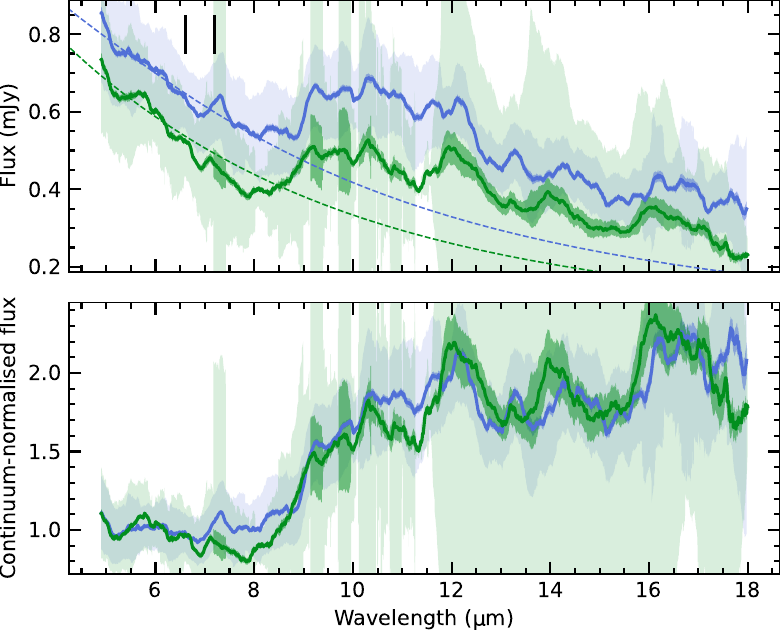}
\caption{\textit{Upper panel}: MIRI MRS spectra for WD\,0145+234, smoothed using a 0.25-{\micron} moving average, with the first epoch shown in blue and the second in green. Thermal continuum models for each epoch are shown with dashed lines, as in Figure~\ref{figureFullspectrum}, with vertical bars marking the tentative carbonates feature. The broad 9--12\,{\micron} feature is consistent with emission from silicate minerals. Shaded regions indicate uncertainties, following Figure~\ref{figureFullspectrum}. \textit{Lower panel}: The same spectra, normalised to the dust thermal continuum by subtracting the stellar blackbody and dividing by the dust blackbody. The broad similarity between the two normalised spectra suggests that the size distribution and mineralogy of the grains producing the silicate feature have not changed dramatically between the two epochs.
\label{figureMIRI}}
\end{figure}

The model closely reproduces the continuum emission from the system at wavelengths shorter than 8\,{\micron}, beyond which the spectrum departs from the blackbody curve due to emission features. Figure~\ref{figureMIRI} shows the MIRI MRS spectrum, where a broad silicate emission feature is clearly present over 9--12\,{\micron}, similar to those at all other white dwarfs observed at these wavelengths \citep{Jura2009silicates}. Several factors influence the strength and morphology of mineralogical emission features, including the size, composition, and environment of the dust grains, so neither their wavelengths nor bandwidths are fixed (e.g.\ \citealt{Posch2007}). Therefore, detailed modelling will be required to characterise the material responsible for the emission, as successfully demonstrated for the prototype dusty white dwarf, G29-38 \citep{Reach2009G29-38, Xu2018IRvariability}. In the meantime, this first-look study offers a qualitative assessment of the data.

There is a subtle feature near 7\,{\micron}, which may arise from carbonate minerals. Inspection of the 2D calibrated images reveals no obvious artefacts that might cause spurious features. However, internal reflections cause fringing in MRS spectra that can exceed one~per cent in amplitude \citep{Argyriou2023, Gasman2023}. Fringe removal is most challenging in the shortest-wavelength channel, which extends to 7.65\,{\micron}. The effect depends on the source position on the detector, so combining multiple dithers may mitigate it. Given the small amplitude of the feature, and the potential for instrumental artefacts in this wavelength region, this feature should be treated with caution until confirmed with fresh data or an improved reduction.

\section{Discussion and outlook}

These observations end the 14-yr hiatus in mid-infrared spectroscopy of white dwarf debris disks. The new data show that small silicate grains appear to be ubiquitous in these disks, and that collisions most likely drive their evolution. Also, \textit{JWST} spectroscopy is nominally sensitive to carbonate minerals, and these may be present among the debris orbiting WD\,0145+234, but require confirmation.

The near-infrared lightcurves shown in Figure\,\ref{figureLightcurve} are broadly consistent with the model proposed to explain the infrared behaviour of the system \citep{Swan2021}. The observational baseline covers thousands of orbital periods, throughout which the flux from the circumstellar debris appears to have closely approximated the $1/t$ decay that is characteristic of a collisional cascade. Notably, the decay appears to be towards the pre-outburst level, rather than the photospheric flux, suggesting that the pre-outburst level represents a long-term minimum. If that is the case, then there may be a constant source of optically thin dust, perhaps via Poynting--Robertson (PR) drag operating on debris previously scattered onto wider orbits \citep{Malamud2021}, or optically thick dust may be present in a stable geometric configuration \citep{Jura2003}. Indeed, while the infrared excesses of dusty white dwarfs are typically variable, none have been observed to fade completely, suggesting that a long-term minimum is a common characteristic of these systems.

The two-blackbody model used to characterise the spectral energy distribution of the star and circumstellar dust is crude but useful, providing a good fit to the continuum below {8\,\micron}. The cooler component representing the dust shows a small but significant increase in temperature between the two MIRI epochs. This variation in the shape of the spectral energy distribution might reflect a radial reconfiguration of the dust, or a change in the grain-size distribution, but a more detailed model is required to explore these ideas. For example, the data should be sufficient to distinguish between different orbital configurations for the circumstellar debris, such as circular or elliptical disks \citep{Dennihy2016}, allowing a more robust determination of the system parameters.

Beyond {8\,\micron}, the observed spectrum departs from the blackbody continuum, showing a prominent emission feature due to silicates. Every white dwarf observed spectroscopically at these wavelengths shows such a feature, although its strength and morphology vary between systems \citep[][figure 6]{farihi2016}. The target shows a relatively modest feature compared to some of its peers, which is interesting as it is the only debris disk with gaseous emission that has been observed with mid-infrared spectroscopy. Systems with such gas features tend to be more variable at 3--{5\,\micron} than other dusty white dwarfs, and are thus inferred to host the most collisionally active disks \citep{Swan2021}, yet it appears that high activity does not necessarily translate to large masses of small, optically thin dust particles.

Only one other white dwarf has mid-infrared spectra taken at multiple epochs (G29-38; \citealt{Xu2018IRvariability}). The silicate features in the two systems have evolved differently, potentially due to their levels of collisional activity. The silicate feature at G29-38 increased in strength in the three years between observations, while its continuum flux measured at {5--7\,\micron} remained stable. By contrast, while the continuum flux at WD\,0145+234 has diminished, its silicate feature has maintained the same relative strength, as shown in Figure~\ref{figureMIRI}. At WD\,0145+234, if the ongoing collisional cascade maintains a constant grain-size distribution, then the strength of the silicate feature would only depend on the mass of emitting dust, as would the continuum flux. At G29-38, if small grains are not being continually replenished in a collisional cascade, they will have been depleted by PR~drag. Therefore, when small grains are later generated in stochastic collisions, they can change the grain-size distribution, and increase the surface area of optically thin dust responsible for the silicate feature.

\begin{figure}
\includegraphics[width=\columnwidth]{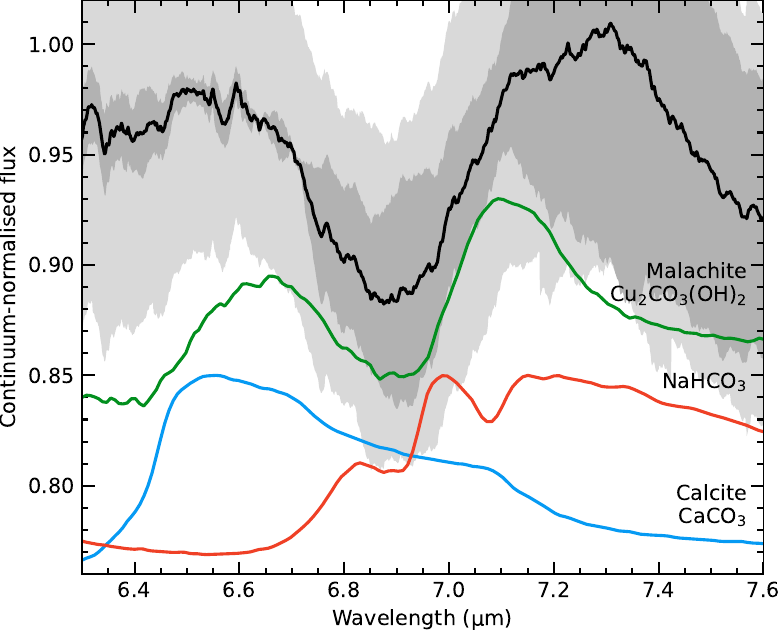}
\caption{Average of the continuum-normalised spectra shown in the bottom panel of Figure~\ref{figureMIRI}, alongside spectra of some carbonate minerals (chosen for their spectral diversity), scaled and offset for visual comparison. Shaded regions indicate uncertainties estimated per-resolution-element (light grey), and by the standard error on the mean (dark grey).
\label{figureCarbonates}}
\end{figure}

It is premature to comment on the detailed mineralogy of the circumstellar debris, as there are numerous outstanding challenges in reducing the data for this faint target (about 0.5\,mJy at {10\,\micron}). Future iterations of the reduction pipeline may improve data quality; therefore, detailed compositional modelling is reserved to later work. Nevertheless, in broad terms, the dominant silicate feature begins near {9\,\micron}, and thus more closely resembles enstatite than forsterite, which are Mg-rich end-members of pyroxene and olivine, respectively. One other spectral feature warrants discussion, namely the subtle peaks near {7\,\micron} that may be due to carbonates. Figure~\ref{figureCarbonates} shows the continuum-normalised MIRI spectrum averaged over both epochs, together with some laboratory spectra of carbonates \citep{Kokaly2017, Christensen2000}. The comparison minerals are selected to demonstrate the potential diversity in spectra of carbonates, where features typically appear near 6.6 and {7.1\,\micron}, but their strength and peak wavelengths vary considerably between minerals. Few other materials exhibit peaks in this region. However, while the data are suggestive, a firm detection cannot be claimed at this stage, so the following discussion is speculative, pending confirmation.

Carbonate minerals are a product of aqueous alteration, produced by reaction with carbon dioxide (as carbonic acid) dissolved in liquid water. If carbonates are indeed present in the dust orbiting WD\,0145+234, that would suggest the parent bodies were large enough (probably tens of km; \citealt{GrimmMcSween1993}) to melt water ices and drive aqueous alteration, consistent with the size of objects whose recent collision is inferred to have caused the infrared outburst \citep{Swan2021}. Solar system objects often display evidence for aqueous alteration. For example, calcite [${\rm CaCO}_{3}$] is common on Earth (e.g.\ as limestone), magnesite [${\rm MgCO}_{3}$] and calcite have been observed on Mars \citep{Bandfield2003,Ehlmann2008, Boynton2009}, and various Na carbonates have been found on the surface of Ceres \citep{Rivkin2006,DeSanctis2016,Carrozzo2018}. Carbonates have also been detected on the small asteroids Ryugu and Bennu \citep{Nakamura2022, Kaplan2020}, which are fragments of bodies similar to those that formed the terrestrial planets. If the first white dwarf planetary system targeted by \textit{JWST} also turns out to host aqueously-altered minerals, that would suggest that liquid water plays a fundamental geochemical role in rocky planet formation beyond the solar system.

The experience gained from this program suggests that \textit{JWST} observations are likely to drive substantial advances in understanding of white dwarf debris disks. The excellent flux calibration of the NIRSPEC instrument will permit more precise characterisation of disk geometry than is possible using \textit{Spitzer} or \textit{WISE} photometry. Separately, the sensitivity of MIRI will allow characterisation of debris disks with much lower fractional luminosities than those probed by \textit{Spitzer}, which represented the brightest members of the population.

The system observed here is one of the brightest dusty white dwarfs, yet is still a challenging target for MIRI MRS. The unpredictable flux variability of white dwarf debris disks complicates matters, as targets may dim substantially between planning and execution of a program. The MIRI Low-Resolution Spectrograph (LRS) is likely to be a better choice than MRS for initial observations, as it has sufficient wavelength coverage (5--{14\,\micron} at $R\sim100$) to reveal both silicates and carbonates, and offers substantial gains in S/N over MRS for the same integration time. Suitable targets for detailed mineralogical studies can be identified using LRS spectra, from which the strength and shape of emission features can be gauged before investing the time required for MRS follow-up.

Observers using MRS should give careful consideration to integration times and the number of dithers required to mitigate the various sources of noise and systematic effects. MIRI photometry may be useful at longer wavelengths, where much higher S/N can be achieved than with spectroscopy. For example, a secondary peak from amorphous silicates appears at {18\,\micron}, which would complement the {10-\micron} feature when fitting models involving compositions and grain sizes. More speculatively, the presence of phyllosilicates (another product of aqueous alteration) can alter the flux distribution between spectral peaks at 21 and {24\,\micron} \citep{Morris2009}, though achieving the required precision to detect that may be challenging, and would require dedicated background observations.

\section*{Acknowledgements}
The authors thank the anonymous referee for a careful and detailed review. AS thanks Boris G\"ansicke for advice and support during the exploratory stages of this work.
This work is based on observations made with the NASA/ESA/CSA \textit{James Webb Space Telescope}. The data were obtained from the Mikulski Archive for Space Telescopes at the Space Telescope Science Institute, which is operated by the Association of Universities for Research in Astronomy, Inc., under NASA contract NAS 5-03127 for JWST. These observations are associated with program \#1647.
This work has received funding from the European Research Council under the European Union’s Horizon 2020 research and innovation programme (Grant agreement No.~101020057), and from NASA through a grant from the Space Telescope Science Institute. KYLS acknowledges support from NASA XRP grant 80NSSC22K0234.

\section*{Data Availability}

The data analysed here can be obtained at \href{https://dx.doi.org/10.17909/q1re-k138}{doi:10.17909/q1re-k138}.



\bibliographystyle{mnras}
\bibliography{WD0145+234_JWST_letter.bib,references_JF,references_KS} 

\begin{thebibliography}{}
\makeatletter
\relax
\def\mn@urlcharsother{\let\do\@makeother \do\$\do\&\do\#\do\^\do\_\do\%\do\~}
\def\mn@doi{\begingroup\mn@urlcharsother \@ifnextchar [ {\mn@doi@}
  {\mn@doi@[]}}
\def\mn@doi@[#1]#2{\def\@tempa{#1}\ifx\@tempa\@empty \href
  {http://dx.doi.org/#2} {doi:#2}\else \href {http://dx.doi.org/#2} {#1}\fi
  \endgroup}
\def\mn@eprint#1#2{\mn@eprint@#1:#2::\@nil}
\def\mn@eprint@arXiv#1{\href {http://arxiv.org/abs/#1} {{\tt arXiv:#1}}}
\def\mn@eprint@dblp#1{\href {http://dblp.uni-trier.de/rec/bibtex/#1.xml}
  {dblp:#1}}
\def\mn@eprint@#1:#2:#3:#4\@nil{\def\@tempa {#1}\def\@tempb {#2}\def\@tempc
  {#3}\ifx \@tempc \@empty \let \@tempc \@tempb \let \@tempb \@tempa \fi \ifx
  \@tempb \@empty \def\@tempb {arXiv}\fi \@ifundefined
  {mn@eprint@\@tempb}{\@tempb:\@tempc}{\expandafter \expandafter \csname
  mn@eprint@\@tempb\endcsname \expandafter{\@tempc}}}

\bibitem[\protect\citeauthoryear{{Alcock}, {Fristrom}  \& {Siegelman}}{{Alcock}
  et~al.}{1986}]{alcock1986}
{Alcock} C.,  {Fristrom} C.~C.,   {Siegelman} R.,  1986, \mn@doi [\apj]
  {10.1086/164005}, \href
  {https://ui.adsabs.harvard.edu/abs/1986ApJ...302..462A} {302, 462}

\bibitem[\protect\citeauthoryear{{Argyriou} et~al.,}{{Argyriou}
  et~al.}{2023a}]{argyriou23_mrs}
{Argyriou} I.,  et~al., 2023a, \mn@doi [\aap] {10.1051/0004-6361/202346489},
  675, A111

\bibitem[\protect\citeauthoryear{Argyriou et~al.,}{Argyriou
  et~al.}{2023b}]{Argyriou2023}
Argyriou I.,  et~al., 2023b, \mn@doi [Astron. Astrophys.]
  {10.1051/0004-6361/202346489}, 675, A111

\bibitem[\protect\citeauthoryear{Argyriou et~al.,}{Argyriou
  et~al.}{2023c}]{argyriou23_brighter_fatter}
Argyriou I.,  et~al., 2023c, \mn@doi [Astron. Astrophys.]
  {10.1051/0004-6361/202346490}, 680, A96

\bibitem[\protect\citeauthoryear{{Bandfield}, {Glotch}  \&
  {Christensen}}{{Bandfield} et~al.}{2003}]{Bandfield2003}
{Bandfield} J.~L.,  {Glotch} T.~D.,   {Christensen} P.~R.,  2003, \mn@doi
  [Science] {10.1126/science.1088054}, \href
  {https://ui.adsabs.harvard.edu/abs/2003Sci...301.1084B} {301, 1084}

\bibitem[\protect\citeauthoryear{{Becklin}, {Farihi}, {Jura}, {Song},
  {Weinberger}  \& {Zuckerman}}{{Becklin} et~al.}{2005}]{becklin2005}
{Becklin} E.~E.,  {Farihi} J.,  {Jura} M.,  {Song} I.,  {Weinberger} A.~J.,
  {Zuckerman} B.,  2005, \mn@doi [\apjl] {10.1086/497826}, \href
  {https://ui.adsabs.harvard.edu/abs/2005ApJ...632L.119B} {632, L119}

\bibitem[\protect\citeauthoryear{B{\"{o}}ker et~al.,}{B{\"{o}}ker
  et~al.}{2023}]{Boker2023}
B{\"{o}}ker T.,  et~al., 2023, \mn@doi [Publ. Astron. Soc. Pacific]
  {10.1088/1538-3873/ACB846}, 135, 038001

\bibitem[\protect\citeauthoryear{{Boynton} et~al.,}{{Boynton}
  et~al.}{2009}]{Boynton2009}
{Boynton} W.~V.,  et~al., 2009, \mn@doi [Science] {10.1126/science.1172768},
  \href {https://ui.adsabs.harvard.edu/abs/2009Sci...325...61B} {325, 61}

\bibitem[\protect\citeauthoryear{{Bushouse} et~al.,}{{Bushouse}
  et~al.}{2023}]{jwstpipeline}
{Bushouse} H.,  et~al., 2023, {JWST Calibration Pipeline}, Zenodo,
  \mn@doi{10.5281/zenodo.7692609}

\bibitem[\protect\citeauthoryear{Carrozzo et~al.,}{Carrozzo
  et~al.}{2018}]{Carrozzo2018}
Carrozzo F.~G.,  et~al., 2018, \mn@doi [Sci. Adv.]
  {10.1126/SCIADV.1701645/SUPPL_FILE/1701645_SM.PDF}, 4

\bibitem[\protect\citeauthoryear{Christensen, Bandfield, Hamilton, Howard,
  Lane, Piatek, Ruff  \& Stefanov}{Christensen et~al.}{2000}]{Christensen2000}
Christensen P.~R.,  Bandfield J.~L.,  Hamilton V.~E.,  Howard D.~A.,  Lane
  M.~D.,  Piatek J.~L.,  Ruff S.~W.,   Stefanov W.~L.,  2000, \mn@doi [J.
  Geophys. Res. Planets] {10.1029/1998JE000624}, 105, 9735

\bibitem[\protect\citeauthoryear{{De Sanctis} et~al.,}{{De Sanctis}
  et~al.}{2016}]{DeSanctis2016}
{De Sanctis} M.~C.,  et~al., 2016, \mn@doi [Nature] {10.1038/nature18290}, 536,
  54

\bibitem[\protect\citeauthoryear{Dennihy, Debes, Dunlap, Dufour, Teske  \&
  Clemens}{Dennihy et~al.}{2016}]{Dennihy2016}
Dennihy E.,  Debes J.~H.,  Dunlap B.~H.,  Dufour P.,  Teske J.~K.,   Clemens
  J.~C.,  2016, \mn@doi [Astrophys. J.] {10.3847/0004-637X/831/1/31}, 831

\bibitem[\protect\citeauthoryear{Dennihy, Clemens, Debes, Dunlap, Kilkenny,
  O'Brien  \& Fuchs}{Dennihy et~al.}{2017}]{Dennihy2017}
Dennihy E.,  Clemens J.~C.,  Debes J.~H.,  Dunlap B.~H.,  Kilkenny D.,  O'Brien
  P.~C.,   Fuchs J.~T.,  2017, \mn@doi [Astrophys. J.]
  {10.3847/1538-4357/aa8ef2}, 849, 77

\bibitem[\protect\citeauthoryear{{Dominik} \& {Decin}}{{Dominik} \&
  {Decin}}{2003}]{Dominik2003}
{Dominik} C.,  {Decin} G.,  2003, \mn@doi [\apj] {10.1086/379169}, \href
  {https://ui.adsabs.harvard.edu/abs/2003ApJ...598..626D} {598, 626}

\bibitem[\protect\citeauthoryear{{Ehlmann} et~al.,}{{Ehlmann}
  et~al.}{2008}]{Ehlmann2008}
{Ehlmann} B.~L.,  et~al., 2008, \mn@doi [Science] {10.1126/science.1164759},
  \href {https://ui.adsabs.harvard.edu/abs/2008Sci...322.1828E} {322, 1828}

\bibitem[\protect\citeauthoryear{{Farihi}}{{Farihi}}{2016}]{farihi2016}
{Farihi} J.,  2016, \mn@doi [\nar] {10.1016/j.newar.2016.03.001}, \href
  {https://ui.adsabs.harvard.edu/abs/2016NewAR..71....9F} {71, 9}

\bibitem[\protect\citeauthoryear{{Farihi}, {Barstow}, {Redfield}, {Dufour}  \&
  {Hambly}}{{Farihi} et~al.}{2010}]{farihi2010}
{Farihi} J.,  {Barstow} M.~A.,  {Redfield} S.,  {Dufour} P.,   {Hambly} N.~C.,
  2010, \mn@doi [\mnras] {10.1111/j.1365-2966.2010.16426.x}, \href
  {https://ui.adsabs.harvard.edu/abs/2010MNRAS.404.2123F} {404, 2123}

\bibitem[\protect\citeauthoryear{{Farihi}, {G{\"a}nsicke}  \&
  {Koester}}{{Farihi} et~al.}{2013}]{farihi2013}
{Farihi} J.,  {G{\"a}nsicke} B.~T.,   {Koester} D.,  2013, \mn@doi [Science]
  {10.1126/science.1239447}, \href
  {https://ui.adsabs.harvard.edu/abs/2013Sci...342..218F} {342, 218}

\bibitem[\protect\citeauthoryear{{Farihi} et~al.,}{{Farihi}
  et~al.}{2022}]{farihi2022}
{Farihi} J.,  et~al., 2022, \mn@doi [\mnras] {10.1093/mnras/stab3475}, \href
  {https://ui.adsabs.harvard.edu/abs/2022MNRAS.511.1647F} {511, 1647}

\bibitem[\protect\citeauthoryear{{Fontaine} \& {Michaud}}{{Fontaine} \&
  {Michaud}}{1979}]{fontaine1979}
{Fontaine} G.,  {Michaud} G.,  1979, \mn@doi [\apj] {10.1086/157247}, \href
  {https://ui.adsabs.harvard.edu/abs/1979ApJ...231..826F} {231, 826}

\bibitem[\protect\citeauthoryear{{G{\"a}nsicke}, {Koester}, {Farihi}, {Parsons}
   \& {Breedt}}{{G{\"a}nsicke} et~al.}{2012}]{gansicke2012}
{G{\"a}nsicke} B.~T.,  {Koester} D.,  {Farihi} J.~{Girven} J.,  {Parsons}
  S.~G.,   {Breedt} E.,  2012, \mn@doi [\mnras]
  {10.1111/j.1365-2966.2012.21201.x}, \href
  {https://ui.adsabs.harvard.edu/abs/2012MNRAS.424..333G} {424, 333}

\bibitem[\protect\citeauthoryear{Gasman et~al.,}{Gasman
  et~al.}{2023}]{Gasman2023}
Gasman D.,  et~al., 2023, \mn@doi [Astron. Astrophys.]
  {10.1051/0004-6361/202245633}, 673, A102

\bibitem[\protect\citeauthoryear{{Gentile Fusillo} et~al.,}{{Gentile Fusillo}
  et~al.}{2021}]{GentileFusillo2021GasDiscs}
{Gentile Fusillo} N.~P.,  et~al., 2021, \mn@doi [Mon. Not. R. Astron. Soc.]
  {10.1093/mnras/stab992}, 504, 2707

\bibitem[\protect\citeauthoryear{{Gianninas}, {Dufour}  \&
  {Bergeron}}{{Gianninas} et~al.}{2004}]{gianninas2004}
{Gianninas} A.,  {Dufour} P.,   {Bergeron} P.,  2004, \mn@doi [\apjl]
  {10.1086/427080}, \href
  {https://ui.adsabs.harvard.edu/abs/2004ApJ...617L..57G} {617, L57}

\bibitem[\protect\citeauthoryear{{Grimm} \& {McSween}}{{Grimm} \&
  {McSween}}{1993}]{GrimmMcSween1993}
{Grimm} R.~E.,  {McSween} H.~Y.,  1993, Science, \href
  {https://ui.adsabs.harvard.edu/abs/1993Sci...259..653G} {259, 653}

\bibitem[\protect\citeauthoryear{{Guidry} et~al.,}{{Guidry}
  et~al.}{2021}]{guidry2021}
{Guidry} J.~A.,  et~al., 2021, \mn@doi [\apj] {10.3847/1538-4357/abee68}, \href
  {https://ui.adsabs.harvard.edu/abs/2021ApJ...912..125G} {912, 125}

\bibitem[\protect\citeauthoryear{{Hollands}, {G{\"a}nsicke}  \&
  {Koester}}{{Hollands} et~al.}{2018}]{hollands2018}
{Hollands} M.~A.,  {G{\"a}nsicke} B.~T.,   {Koester} D.,  2018, \mn@doi
  [\mnras] {10.1093/mnras/sty592}, \href
  {https://ui.adsabs.harvard.edu/abs/2018MNRAS.477...93H} {477, 93}

\bibitem[\protect\citeauthoryear{{Jakobsen} et~al.,}{{Jakobsen}
  et~al.}{2022}]{jakobsen2022}
{Jakobsen} P.,  et~al., 2022, \mn@doi [\aap] {10.1051/0004-6361/202142663},
  \href {https://ui.adsabs.harvard.edu/abs/2022A&A...661A..80J} {661, A80}

\bibitem[\protect\citeauthoryear{Jura}{Jura}{2003}]{Jura2003}
Jura M.,  2003, \mn@doi [Astrophys. J. Lett.] {10.1086/374036}, 584, L91

\bibitem[\protect\citeauthoryear{{Jura} \& {Young}}{{Jura} \&
  {Young}}{2014}]{jura2014}
{Jura} M.,  {Young} E.~D.,  2014, \mn@doi [Annual Review of Earth and Planetary
  Sciences] {10.1146/annurev-earth-060313-054740}, \href
  {https://ui.adsabs.harvard.edu/abs/2014AREPS..42...45J} {42, 45}

\bibitem[\protect\citeauthoryear{Jura, Farihi  \& Zuckerman}{Jura
  et~al.}{2009}]{Jura2009silicates}
Jura M.,  Farihi J.,   Zuckerman B.,  2009, \mn@doi [Astron. J.]
  {10.1088/0004-6256/137/2/3191}, 137, 3191

\bibitem[\protect\citeauthoryear{Kaplan et~al.,}{Kaplan
  et~al.}{2020}]{Kaplan2020}
Kaplan H.~H.,  et~al., 2020, \mn@doi [Science (80-. ).]
  {10.1126/SCIENCE.ABC3557/SUPPL_FILE/ABC3557_KAPLAN_SM.PDF}, 370, eabc3557

\bibitem[\protect\citeauthoryear{Kenyon \& Bromley}{Kenyon \&
  Bromley}{2017}]{Kenyon2017collisions}
Kenyon S.~J.,  Bromley B.~C.,  2017, \mn@doi [Astrophys. J.]
  {10.3847/1538-4357/aa7b85}, 844, 116

\bibitem[\protect\citeauthoryear{{Kilic}, {von Hippel}, {Leggett}  \&
  {Winget}}{{Kilic} et~al.}{2005}]{kilic2005}
{Kilic} M.,  {von Hippel} T.,  {Leggett} S.~K.,   {Winget} D.~E.,  2005,
  \mn@doi [\apjl] {10.1086/497825}, \href
  {https://ui.adsabs.harvard.edu/abs/2005ApJ...632L.115K} {632, L115}

\bibitem[\protect\citeauthoryear{{Klein}, {Jura}, {Koester}, {Zuckerman}  \&
  {Melis}}{{Klein} et~al.}{2010}]{klein2010}
{Klein} B.,  {Jura} M.,  {Koester} D.,  {Zuckerman} B.,   {Melis} C.,  2010,
  \mn@doi [\apj] {10.1088/0004-637X/709/2/950}, \href
  {https://ui.adsabs.harvard.edu/abs/2010ApJ...709..950K} {709, 950}

\bibitem[\protect\citeauthoryear{{Koester}, {Provencal}  \&
  {Shipman}}{{Koester} et~al.}{1997}]{koester1997}
{Koester} D.,  {Provencal} J.,   {Shipman} H.~L.,  1997, \aap, \href
  {https://ui.adsabs.harvard.edu/abs/1997A&A...320L..57K} {320, L57}

\bibitem[\protect\citeauthoryear{Kokaly et~al.,}{Kokaly
  et~al.}{2017}]{Kokaly2017}
Kokaly R.~F.,  et~al., 2017, \mn@doi [U.S. Geol. Surv. Data Ser.]
  {10.3133/DS1035}, 1035, 61

\bibitem[\protect\citeauthoryear{Malamud, Grishin  \& Brouwers}{Malamud
  et~al.}{2021}]{Malamud2021}
Malamud U.,  Grishin E.,   Brouwers M.,  2021, \mn@doi [Mon. Not. R. Astron.
  Soc.] {10.1093/mnras/staa3940}, 501, 3806

\bibitem[\protect\citeauthoryear{McCleery et~al.,}{McCleery
  et~al.}{2020}]{McCleery2020}
McCleery J.,  et~al., 2020, \mn@doi [Mon. Not. R. Astron. Soc.]
  {10.1093/mnras/staa2030}, 499, 1890

\bibitem[\protect\citeauthoryear{Melis, Klein, Doyle, Weinberger, Zuckerman  \&
  Dufour}{Melis et~al.}{2020}]{Melis2020gas}
Melis C.,  Klein B.,  Doyle A.~E.,  Weinberger A.~J.,  Zuckerman B.,   Dufour
  P.,  2020, \mn@doi [Astrophys. J.] {10.3847/1538-4357/abbdfa}, 905, 56

\bibitem[\protect\citeauthoryear{Morris \& Desch}{Morris \&
  Desch}{2009}]{Morris2009}
Morris M.~A.,  Desch S.~J.,  2009, \mn@doi [Astrobiology]
  {10.1089/AST.2008.0316}, 9, 965

\bibitem[\protect\citeauthoryear{{Nakamura} et~al.,}{{Nakamura}
  et~al.}{2022}]{Nakamura2022}
{Nakamura} E.,  et~al., 2022, \mn@doi [Proceedings of the Japan Academy, Series
  B] {10.2183/pjab.98.015}, \href
  {https://ui.adsabs.harvard.edu/abs/2022PJAB...98..227N} {98, 227}

\bibitem[\protect\citeauthoryear{Posch, Baier, Mutschke  \& Henning}{Posch
  et~al.}{2007}]{Posch2007}
Posch T.,  Baier A.,  Mutschke H.,   Henning T.,  2007, \mn@doi [Astrophys. J.]
  {10.1086/521390/FULLTEXT/}, 668, 993

\bibitem[\protect\citeauthoryear{Reach, Lisse, von Hippel, Mullally, von Hippel
   \& Mullally}{Reach et~al.}{2009}]{Reach2009G29-38}
Reach W.~T.,  Lisse C.,  von Hippel T.,  Mullally F.,  von Hippel T.,
  Mullally F.,  2009, \mn@doi [Astrophys. J.] {10.1088/0004-637X/693/1/697},
  693, 697

\bibitem[\protect\citeauthoryear{{Rieke} et~al.,}{{Rieke}
  et~al.}{2015}]{rieke2015}
{Rieke} G.~H.,  et~al., 2015, \mn@doi [\pasp] {10.1086/682252}, \href
  {https://ui.adsabs.harvard.edu/abs/2015PASP..127..584R} {127, 584}

\bibitem[\protect\citeauthoryear{{Rivkin}, {Volquardsen}  \& {Clark}}{{Rivkin}
  et~al.}{2006}]{Rivkin2006}
{Rivkin} A.~S.,  {Volquardsen} E.~L.,   {Clark} B.~E.,  2006, \mn@doi [\icarus]
  {10.1016/j.icarus.2006.08.022}, \href
  {https://ui.adsabs.harvard.edu/abs/2006Icar..185..563R} {185, 563}

\bibitem[\protect\citeauthoryear{Rocchetto, Farihi, G{\"{a}}nsicke  \&
  Bergfors}{Rocchetto et~al.}{2015}]{Rocchetto2015}
Rocchetto M.,  Farihi J.,  G{\"{a}}nsicke B.~T.,   Bergfors C.,  2015, \mn@doi
  [Mon. Not. R. Astron. Soc.] {10.1093/mnras/stv282}, 449, 574

\bibitem[\protect\citeauthoryear{{Schatzman}}{{Schatzman}}{1958}]{schatzman1958}
{Schatzman} E.~L.,  1958, {White dwarfs}.
North-Holland Pub Co

\bibitem[\protect\citeauthoryear{{Swan}, {Farihi}, {Koester}, {Holls},
  {Parsons}, {Cauley}, {Redfield}  \& {G{\"a}nsicke}}{{Swan}
  et~al.}{2019}]{swan2019}
{Swan} A.,  {Farihi} J.,  {Koester} D.,  {Holls} M.,  {Parsons} S.,  {Cauley}
  P.~W.,  {Redfield} S.,   {G{\"a}nsicke} B.~T.,  2019, \mn@doi [\mnras]
  {10.1093/mnras/stz2337}, \href
  {https://ui.adsabs.harvard.edu/abs/2019MNRAS.490..202S} {490, 202}

\bibitem[\protect\citeauthoryear{Swan, Farihi, Wilson  \& Parsons}{Swan
  et~al.}{2020a}]{Swan2020dust}
Swan A.,  Farihi J.,  Wilson T.~G.,   Parsons S.~G.,  2020a, \mn@doi [Mon. Not.
  R. Astron. Soc.] {10.1093/mnras/staa1688}, 496, 5233

\bibitem[\protect\citeauthoryear{{Swan}, {Farihi}  \& {Wilson}}{{Swan}
  et~al.}{2020b}]{swan2020}
{Swan} A.,  {Farihi} J.,   {Wilson} Thomas G.~{Parsons} S.~G.,  2020b, \mn@doi
  [\mnras] {10.1093/mnras/staa1688}, \href
  {https://ui.adsabs.harvard.edu/abs/2020MNRAS.496.5233S} {496, 5233}

\bibitem[\protect\citeauthoryear{Swan, Kenyon, Farihi, Dennihy, G{\"{a}}nsicke,
  Hermes, Melis  \& von Hippel}{Swan et~al.}{2021}]{Swan2021}
Swan A.,  Kenyon S.~J.,  Farihi J.,  Dennihy E.,  G{\"{a}}nsicke B.~T.,  Hermes
  J.~J.,  Melis C.,   von Hippel T.,  2021, \mn@doi [Mon. Not. R. Astron. Soc.]
  {10.1093/MNRAS/STAB1738}, 506, 432

\bibitem[\protect\citeauthoryear{{Vanderbosch} et~al.,}{{Vanderbosch}
  et~al.}{2021}]{vanderbosch2021}
{Vanderbosch} Z.~P.,  et~al., 2021, \mn@doi [\apj] {10.3847/1538-4357/ac0822},
  \href {https://ui.adsabs.harvard.edu/abs/2021ApJ...917...41V} {917, 41}

\bibitem[\protect\citeauthoryear{{Vanderburg} et~al.,}{{Vanderburg}
  et~al.}{2015}]{vanderburg2015}
{Vanderburg} A.,  et~al., 2015, \mn@doi [\nat] {10.1038/nature15527}, \href
  {https://ui.adsabs.harvard.edu/abs/2015Natur.526..546V} {526, 546}

\bibitem[\protect\citeauthoryear{{\noopsort{Vanmaanen}}van~Maanen}{{\noopsort{Vanmaanen}}van~Maanen}{1917}]{vanmaanen1917}
{\noopsort{Vanmaanen}}van~Maanen A.,  1917, \mn@doi [\pasp] {10.1086/122654},
  \href {https://ui.adsabs.harvard.edu/abs/1917PASP...29..258V} {29, 258}

\bibitem[\protect\citeauthoryear{{Wang} et~al.,}{{Wang}
  et~al.}{2019}]{wang2019}
{Wang} T.-g.,  et~al., 2019, \mn@doi [\apjl] {10.3847/2041-8213/ab53ed}, \href
  {https://ui.adsabs.harvard.edu/abs/2019ApJ...886L...5W} {886, L5}

\bibitem[\protect\citeauthoryear{{Wells} et~al.,}{{Wells}
  et~al.}{2015}]{Wells2015}
{Wells} M.,  et~al., 2015, \mn@doi [\pasp] {10.1086/682281}, \href
  {https://ui.adsabs.harvard.edu/abs/2015PASP..127..646W} {127, 646}

\bibitem[\protect\citeauthoryear{{Wyatt}, {Farihi}, {Pringle}  \&
  {Bonsor}}{{Wyatt} et~al.}{2014}]{wyatt2014}
{Wyatt} M.~C.,  {Farihi} J.,  {Pringle} J.~E.,   {Bonsor} A.,  2014, \mn@doi
  [\mnras] {10.1093/mnras/stu183}, \href
  {https://ui.adsabs.harvard.edu/abs/2014MNRAS.439.3371W} {439, 3371}

\bibitem[\protect\citeauthoryear{{Xu}, {Zuckerman}, {Dufour}, {Young}, {Klein}
  \& {Jura}}{{Xu} et~al.}{2017}]{xu2017}
{Xu} S.,  {Zuckerman} B.,  {Dufour} P.,  {Young} E.~D.,  {Klein} B.,   {Jura}
  M.,  2017, \mn@doi [\apjl] {10.3847/2041-8213/836/1/L7}, \href
  {https://ui.adsabs.harvard.edu/abs/2017ApJ...836L...7X} {836, L7}

\bibitem[\protect\citeauthoryear{Xu et~al.,}{Xu
  et~al.}{2018}]{Xu2018IRvariability}
Xu S.,  et~al., 2018, \mn@doi [Astrophys. J.] {10.3847/1538-4357/aadcfe}, 866,
  108

\bibitem[\protect\citeauthoryear{{Zuckerman} \& {Becklin}}{{Zuckerman} \&
  {Becklin}}{1987}]{zuckerman1987}
{Zuckerman} B.,  {Becklin} E.~E.,  1987, \mn@doi [\nat] {10.1038/330138a0},
  \href {https://ui.adsabs.harvard.edu/abs/1987Natur.330..138Z} {330, 138}

\bibitem[\protect\citeauthoryear{{Zuckerman}, {Koester}, {Melis}, {Hansen}  \&
  {Jura}}{{Zuckerman} et~al.}{2007}]{zuckerman2007}
{Zuckerman} B.,  {Koester} D.,  {Melis} C.,  {Hansen} B.~M.,   {Jura} M.,
  2007, \mn@doi [\apj] {10.1086/522223}, \href
  {https://ui.adsabs.harvard.edu/abs/2007ApJ...671..872Z} {671, 872}

\makeatother
\end{thebibliography}








\bsp	
\label{lastpage}
\end{document}